\documentclass[onecolumn]{aastex}
\usepackage{spr-astr-addons}
\usepackage{url}

\RequirePackage{color}

\newcommand{\emaila}{}

\def\bea{\begin{eqnarray}}
\def\ena{\end{eqnarray}}

\begin{document}
\title{Radio precursors to neutron star binary mergings}
\shorttitle{Radio transients from NS binary mergings}

\author{M. S. Pshirkov\altaffilmark{1}} \and \author{K. A. Postnov\altaffilmark{2}}


\emaila{pshirkov@prao.ru}

\altaffiltext{1}{Pushchino Radio Astronomy Observatory, 142290 Pushchino, Russia}
\altaffiltext{2}{Sternberg Astronomical Institute, 119992 Moscow, Russia}

\begin{abstract}
We discuss a possible generation of  radio bursts
preceding final stages of binary neutron star mergings which can be
accompanied by short gamma-ray bursts.
Detection of such bursts appear to be advantageous in the low-frequency radio
band due to a time delay of ten to several hundred seconds required for
radio signal to propagate in the ionized intergalactic medium.
This delay makes it possible to use short gamma-ray burst alerts to
promptly monitor specific regions on the sky by
low-frequency radio facilities, especially by LOFAR.
To estimate the strength of the radio signal, we assume a power-law dependence of the radio luminosity
on the total energy release in a magnetically dominated outflow,
as found in millisecond pulsars. Based on the planned
LOFAR sensitivity at 120 MHz, we estimate that the LOFAR detection rate of such
radio transients could be about several events per month from redshifts up to $z\sim1.3$ in the most optimistic scenario.
The LOFAR ability to detect such events would crucially  depend on exact efficiency of low-frequency radio emission mechanism.
\end{abstract}



\section{Introduction}
\label{SectionI}

Despite the decade of active researches, cosmic
gamma-ray bursts (GRBs) remain in the focus of modern astrophysical
studies. A huge electromagnetic energy output of $\sim 10^{48}-10^{53}$~ergs
observed in GRBs requires gravitational or rotational energy release
possibly mediated by magnetic field
under extreme conditions (e.g., core collapse of massive rotating stars,
binary neutron star (NS) or NS -- black hole (BH) binary mergings).
Among many possibilities, the concept of collapsar \citep{Woosley1993}
for long GRBs (LGRBs) and compact binary mergings for short GRBs (SGRBs)
(e.g. \cite{Blinnikov1984,Eichler1989}, see a review by \cite{Nakar2007}) appear
to be the most viable ones. However, we are apparently still far from
full understanding of these most energetic transient natural
phenomena (see, e.g., the recent critical discussion by \cite{Lyutikov2009}).

A magnetic mechanism may be required to explain the rich phenomenology
of GRBs (e.g. \cite{Barkov2008}). A BH surrounded by magnetized torus
seems to be the prerequisite condition to form a GRB, since collimated relativistic
outflows can not be produced by electromagnetic mechanism without external pressure
\citep{Lyubarsky2009}.

It is very challenging to seek for various
messengers from complicated physical processes involved,
such as gravitational-wave bursts
\citep{Sengupta2009} expected from compact binary mergings,
active neutrino emission \citep{Ruffert1997} and
afterglows in the broad range of electromagnetic frequencies from
radio  and optics
to x-ray \citep{Kann2008, Nysewander2009}.
The GRB afterglows are associated with the interaction of the
relativistic GRB ejecta with the surrounding medium (see e.g. \cite{Hurley2006} for a review) and
will not be considered here.

In the hard electromagnetic domain, the so-called GRB precursors preceding
the main burst are found for a sizeable fraction of long GRBs with spectral properties similar to the main GRB emission (see
\cite{Burlon2008}), but none has been detected so far for short GRBs.
Yet there are prospects for radio
precursors for the merging of magnetized binary NS as well
(e.g. \citep{Lipunov1996,  Hansen2001, Moortgat2005}).
If low-frequency radio emission can be generated prior to a
SGRB, due to the
dispersion  in the intergalactic plasma the radio signal would arrive \textit{later}
than the gamma-ray pulse \citep{Lipunov_ea97}. So the GRB itself
may be used as a trigger to search for such radio transient.

In this paper we would like to consider yet another possibility
of the formation of non-thermal low-frequency radio emission that
can be
generated in relativistic plasma outflow prior to the final collapse of a binary neutron star. In contrast to the previous work by Hansen \& Lyutikov,
in which magnetospheric pair plasma generation was considered before the
destruction of merging neutron stars, we shall investigate the next phase of
the merging when a single differentially rotating object is formed and when
the magnetic field can be amplified to magnetar values.

Our motivations for focusing
on low-frequency radio emission are twofold.  First,
the lower the frequency, the longer is the intergalactic delay and
the more time is accessible for a radio telescope to point to the
GRB position.  At frequencies above 1~GHz
the delay is about a few seconds, which is insufficient
for repointing of a large dish. The second motivation is stimulated by
the approaching start of operation  of the LOFAR radio telescope, which will
have a record high sensitivity in the low-frequency range and whose design
allows one to react to alerts with the necessary rapidity  \citep{Fenderetal2008,vLeeuwen2009a,vLeeuwen2009b}.

Our further considerations are based on the following three assumptions: (i)
SGRBs are produced by binary NS mergers; (ii) an ultra-strong
magnetic field ($\sim10^{15-16}~\mathrm{G}$) is needed to power
the GRB engine; (iii) a rapidly rotating pre-GRB object with
strong magnetic field which is formed immediately after the
merging can radiate in radio waves very much the
same as usual pulsar.

 Neutron stars are known to have strong
magnetic fields that can survive through long time of the NS life
(e.g. internal toroidal field, \cite{Abdolrahimi2009}). It is quite
natural that these fields could give rise to various
electromagnetic phenomena during final stages of the binary NS
coalescence.

Binary NS population with the time prior to the coalescence due to
gravitational wave emission less than the age of the Universe is
known from binary pulsar observations \citep{Regimbau2005}. The
rate of binary NS mergings is estimated to be quite high,
$\mathcal{R}_{NS}\sim 10^{2}-10^3~\rm Gpc^{-3}yr^{-1}$ (see
\cite{PostnovYungelson06} for a review), which is about two order
of magnitudes higher than the estimated rate of SGRBs:
$\mathcal{R}_{SGRBs}\sim 1-10~\rm Gpc^{-3}yr^{-1}$
\citep{Nakar2007}. This discrepancy, of course, is not dangerous
for binary NS mergings as the SGRB model considering narrow
gamma-ray beaming and the possibility that not every merging ends
up with a burst because of lack of required physical conditions
(insufficient magnetic field, low mass, etc.).

Bursting electromagnetic emission can be generated  at different stages of the
merging process. First, a joint magnetosphere of two coalescing
NSs can be restructured to produce strong flares from reconnection
of magnetic lines (this process should be considered separately
and will be discussed elsewhere). Next, during several last
orbital revolutions before the merging the magnetospheric plasma
effects come into play (e.g. \cite{Lipunov1996,Hansen2001}).
Simulations show that after the merging  a massive object with
rapid differential rotation is formed \citep{Shibata2005,
Rasio2005,Faber2006}, see review by \cite{Duez2009}. This object can not immediately collapse
into BH
until it gets rid of excessive  angular
momentum via some mechanism. At
this stage a significant increase of the seed magnetic fields of
NSs can occur: the energy of the differential rotation can be
effectively transformed into the energy of magnetic field. Here
the situation may be similar to what is thought to occur during
the NS formation in the core collapse supernovae \citep{Spruit2008}.

The growing magnetic field and rapid rotation can lead to the
relativistic plasma generation and the formation of the outflow
along the open magnetic field lines, in which pulsar-like
low-frequency radiation can be produced.
As we shall see, such a rapidly rotating strongly
magnetized object can have
much higher radio luminosity than even the brightest ordinary
pulsars. Finally, the merging remnant can collapse into a BH
(provided that its mass is above the maximum mass of a NS),
possibly surrounded by a magnetized torus; such a configuration
is favorable for the launch of a GRB.


\section{Magnetic field amplification \label{SectionII}}

First we address the question of the magnetic field
amplification after the merging which is absolutely necessary for  significant
radio-emission generation.

Full MHD-simulations of the merging process in GR
are extremely complicated and have not been  performed as yet \citep{Duez2009}, so
we have to  use  crude semi-qualitative estimates.

The magnetic field amplification in the differentially rotating configuration
occurs at the expense of the energy contained in the differential rotation and
can be estimated as  \citep{Spruit2008}:
\bea
B^2R^3\sim\left(\frac{\Delta \Omega}{\Omega}\right)^2\Delta
E, \label{limits_mf_1}
\ena
where $R$ is the characteristic radius
of the region occupied by the strong magnetic field, $R\approx 10^{6}~\rm cm$ in our case,
$\frac{\Delta\Omega}{\Omega}$ is the factor characterizing
the differential rotation and $\Delta E$ is the full
rotational energy. For binary NS mergings we expect
$$\Delta
E\sim E_{\rm orb}\sim 10^{53}~\rm erg\,.
$$
This estimate shows that the magnetic field amplified during the binary NS coalescence
can in principle be by an order
of magnitude higher than the NS magnetic field generated in stellar core collapses:
\bea
\frac{B_{\rm coal}^2}{B_{\rm SN}^2}\sim \frac{\Delta E_{\rm
coal}}{\Delta E_{\rm SN}}\sim\frac{10^{53}~\rm erg}{10^{51}~\rm
erg}\sim 100 , \label{limits_mf_2}
\ena
Taking as granted a
magnetar field ($\sim 10^{15}~\mathrm{G}$), we arrive at a fiducial value of $B_{max} =
10^{16}~\rm G$ during the NS merging.
These considerations are backed with recent numerical simulations
\citep{Duez2006}. Some models predict even larger
values of the resulting field (\cite{Usov1992}, up to $10^{17}$ G),
but for conservative estimates we shall not use them.

The amplification of the  poloidal field in the differentially rotating
post-merger configuration may occur linearly (due to magnetic winding) or
exponentially (due to magneto-rotational instability) \citep{Duez2006}.
The actual value of the maximal field attained depends on time available
in the differentially rotating configuration: e.g., it can be destroyed by vigorous emission of gravitational waves in
case if it was not precisely axial symmetric; such bar-like
configurations may emerge in case of stiff EoS
\citep{Shibata2005}. Also, if it can
survive long enough, the field can be  amplified to the limits
imposed by some external factor, e.g. the hydrostatic magnetic
buoyancy \citep{Kluzniak1998} -- a toroidal configuration with
strong magnetic field can float to the surface, thus effectively
stopping further enhancement of the field.

The energy loss rate in a magnetically driven outflow $\dot{E}_m$ depends on the
magnitude of the amplified field $B$, the characteristic
angular rotation frequency $\Omega$ and the size of the object $R$:
\bea
\dot{E}_{\rm m}\sim\frac{\Omega^4B^2R^6}{c^3}
\label{grb_luminosity}
\ena
To be associated
with the electromagnetic luminosity of GRBs
$\dot{E}_{\rm GRB}=\dot E_m$, for typical values $R\sim 10^6$~cm and
$\Omega\sim 4000-6000$ the magnetic field must fall within the range
\bea
B\sim (10^{14}-10^{16}) ~\rm G\,.
\label{mf_from_grb}
\ena
This estimate is in agreement with the expected field duering the merging process.

Numerical simulations of binary NS mergings by \cite{Kiuchi2009} suggest
the time prior to collapse of order of a few ten ms almost independently
on the initial conditions.
Assuming the field amplification time to be the same for
all SGRBs, the final distribution of magnetic fields should follow the initial one.
This statement seems to be implicitly confirmed by the similarity between the observed
SGRBs luminosity function and the distribution of magnetic field in coalescing
binary NS prior to the merging \citep{Postnov2009}.

\section{Observations of rapidly rotating magnetar  with LOFAR}
\label{SectionIII}

A rapidly rotating post-merger object with strong magnetic field may  generate low-frequency radio emission by the same physical
mechanism as in ordinary pulsars.
The exact mechanism of production of pulsar radio emission is a matter of debates
(see e.g. \cite{Lyubarsky2008} for a review), so we will treat the problem phenomenologically and assume that the radio luminosity is proportional
to the total rotational energy losses
\begin{equation}
L_{\mathrm rad}=\eta\dot{E},
\label{radio_energy}
\end{equation}

\begin{figure}
\center
\includegraphics[width=86 mm]{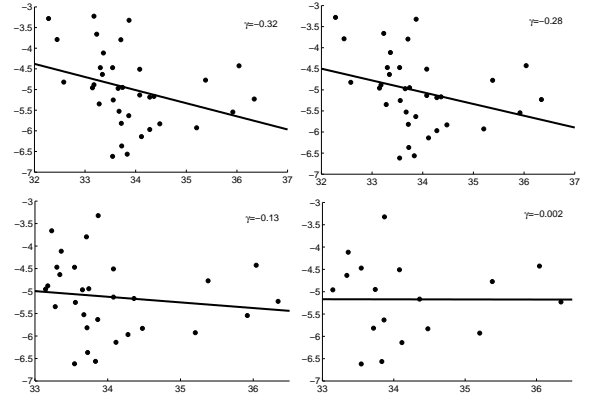}
\caption{
Conversion efficiency $\eta$ of the spin-down power $\dot{E}$ into radio emission
as a function $\dot{E}$ (log-log)  for (a) all millisecond pulsars (upper left), (b) millisecond pulsars with spin frequencies $f>$ 50~Hz (upper right), (c) 100~Hz (bottom left) and (d) 200 Hz~(bottom right). $\eta$ is fitted by the power-law $\eta \propto{\dot{E}^{\gamma}}$.
For rapidly rotating millisecond pulsars the conversion efficiency
becomes weakly dependent on $\dot{E}$. The parameters of MSPs were taken from \cite{atnfpulsars}.}
\label{eta_msp}
\end{figure}

\noindent with the conversion coefficient $\eta$ being a function of
the total energy loss rate $\dot{E}$.
Using plots shown in Figure \ref{eta_msp}, for further estimates
we adopted the power-law dependence
\begin{equation}
\eta(\dot{E})=10^{-5}\left(\frac{\dot{E}}{10^{35}~\mathrm{erg~s^{-1}}}\right)^{\gamma},
\label{eta_dependency}
\end{equation}
with $-1/2<\gamma<0$.
The lower bound -1/2 (i.e. $\eta\sim 1/\sqrt{\dot E}$)
can be derived from the efficiency of the secondary
pair production mechanism in ordinary pulsars \cite{Lyubarsky2008}.

For estimates of the emerging radio flux, we adopt a power-law radio spectrum
with index $\alpha =-2$:
$F(f)\propto f^{-2}$.
In our estimates we assume $\dot{E}=10^{50\div 52}~\mathrm{erg~s^{-1}}$.
Then the expected radio flux density at the fiducial
lower cut-off frequency 100 MHz will be
\bea
F \sim 8\cdot 10^{3+15\gamma} \dot{E}_{50}^{1+\gamma}
\left(\frac{1~\mathrm{Gpc}}{D}\right)^2 \mathrm{Jy}\,.
\label{flux_density_1}
\ena
where $\dot{E}_{50}=\dot{E}/10^{50} \mathrm{erg s^{-1}}$ .

The duration of the radio pulse will be shorter than the time between the
merging and the final collapse to BH, i.e. a few tens of ms\footnote{It is possible, however, that the duration can
be extended up to a few seconds in some models \citep{Duez2009}. In that case the observed radio signal would be
two-three orders of magnitude brighter.}. It is not excluded
that the enigmatic ms radio transient
reported by \cite{Lorimer2007} actually originated from the binary NS coalescence event.

Can such a bright short radio
transient event be detected by LOFAR?
Bearing in mind the assumed power-law spectrum
we are primarily interested in the 120 MHz band of the High-Band Antenna (HBA) of
LOFAR.
Low-frequency radio observations are hampered with effects of propagation in the interstellar and intergalactic medium, especially due to strong dispersion and time scattering. The first effect can be treated by means of de-dispersion, at the same time providing convincing argument about extragalactic origin of the transient; moreover,
only time delay due to the interstellar/intergalactic
dispersion makes the observation of the
precursor \textit{after} GRB-alert possible. On the contrary,  the interstellar/intergalactic scattering has purely devastating effect on the observations, irreversibly smearing short pulses. This broadening very strongly depends on the frequency of observations \citep{Bhatetal2004}
$\tau_{\mathrm sc} \propto f^{-4}$.
So a pulse with intrinsic width $\sim10~\mathrm{ms}$ and flux density (\ref{flux_density_1}) will be broadened to $\tau_{\mathrm{sc}}\sim10^2(D/1~\mathrm{Gpc})^2$~s and the resulting flux density will be reduced:
\bea
F_{\mathrm{obs}}(120~ \mathrm{MHz}) \sim 6\cdot
10^{2+15\gamma} \dot{E}_{50}^{1+\gamma}\left(\frac{1~\mathrm{Gpc}}{D}\right)^4~ \mathrm{mJy}\,
\label{flux_density_2}
\ena

Using LOFAR parameters from \cite{Nijboer2009},
we obtain for the configuration with  13 core + 7 remote stations the following sensitivity:
\bea
S_{\mathrm 13+7} = 40 \left(\frac{SNR}{10}\right)\left(\frac{D}{1~\mathrm{Gpc}}\right)^{-1}\left(\frac{\Delta f}{4~\mathrm{MHz}}\right)^{-1/2}~\mathrm{mJy},
\label{sensitivity_lofar_1}
\ena
where $\Delta f$ is the bandwidth.
The maximal  distance $D$ from which radio precursor can be detected by LOFAR gives us
the rate of SGRBs with the assumed total energy release:
 \begin{equation}
 D=\left(1.5\cdot10^{1+15\gamma}\dot{E}_{50}^{1+\gamma}\right)^{1/3}~\mathrm{Gpc}
 \label{Dmax}
 \end{equation}


\begin{figure}
\center
\includegraphics[width=86 mm]{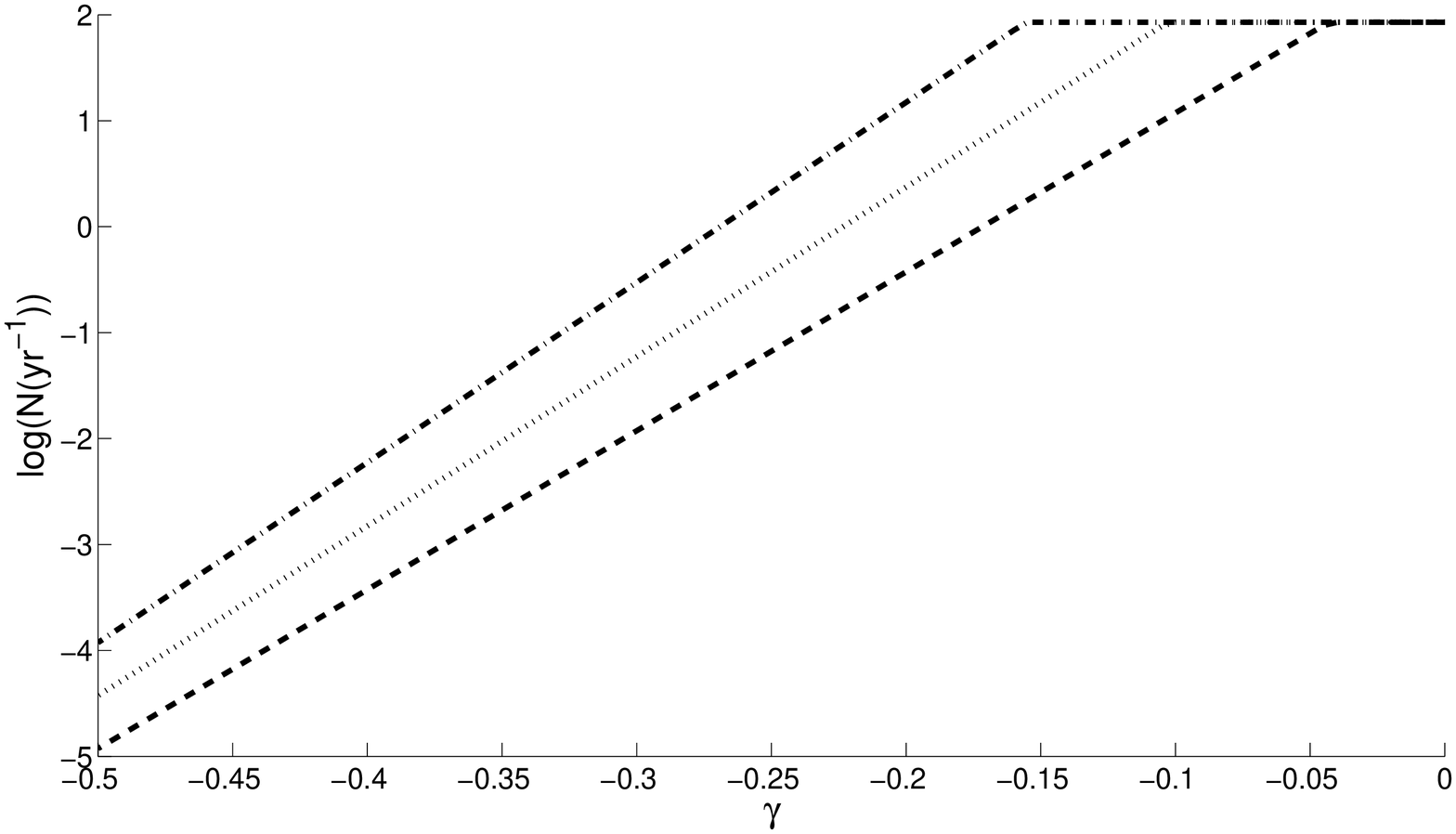}
\caption{
The expected rate of LOFAR detections as a function of the slope parameter $\gamma$ (see eq. (\ref{eta_dependency})) for various total energy loss rates: the dash, dotted, and dash-dotted lines are shown for $E_{50}=1$, 10, and 100, respectively.}
\label{gamma_rate}
\end{figure}

According to the BATSE catalog, the SGRB rate  is about 170 per year \citep{Meegan1997}, around 30 per cent of them are found
at redshifts smaller than $z=0.2$ ($D=1$ Gpc) \citep{Guetta2006}; we have used these figures to obtain the expected rate of LOFAR detections shown in Fig. \ref{gamma_rate}.

\section{Conclusions}
\label{SectionIV}
A binary NS coalescence leads to the formation of
a differentially rotating massive object which can eventually collapse to
BH possibly surrounded by a magnetized torus. Numerical simulations show \citep{Duez2006} that
the magnetic field is strongly amplified at the pre-collapse stage
up to $10^{15}-10^{16}$~G. This configuration is
favorable for the formation of
a short gamma-ray burst.

At the stage preceding the collapse and GRB, a relativistic plasma outflow
can be produced by the differentially rotating strongly magnetized configuration.
Some fraction of the
total power of this outflow can be converted to
electromagnetic radio emission.
The main uncertainty in the estimation of
the radio luminosity before the collapse of this configuration
is the unknown efficiency $\eta$ of the conversion of the rotational energy
losses from the differentially rotating pre-collapse object
into radio emission, which we assumed to depend on
the total energy loss rate as $\eta\propto(\dot{E})^{\gamma}$ in analogy with millisecond
radio pulsars.
We have shown that  in the most optimistic case the planned LOFAR sensitivity at 120 MHz
allows the potential detection at the signal-to-noise ratio level $>10$ of short
(the proper duration of order of 10-100 ms, smeared by scattering
in the intergalactic medium to $\sim 100$ s)
radio bursts associated with
SGRBs from redshifts $z<1.3$ at a rate of $\sim$ 90 events/year.
For an assumed dispersion measure of 1000 cm$^{-3}$pc,
the low-frequency 120 MHz radio bursts should be delayed by 290 s
with respect to the prompt gamma-ray emission, which enables one to use
SGRB alerts for rapid redirecting the LOFAR synthesized beam to the SGRB position on the sky.

The detection of non-thernal radio bursts associated with short GRBs
will strongly indicate the involvement of high magnetic field in the GRB engine.
The dispersion measure of such a burst obtained from radio observations
will give an independent direct estimate of the GRB distance.
Radio precursors to short GRBs detected by LOFAR will
open up an interesting
possibility to search for such transients in LOFAR all-sky surveys. Short radio transients
with similar characteristics
can be a signature of 'orphan' SGRBs.
This information can also be used in the analysis of data obtained by
modern gravitational wave detectors.



\section*{Acknowledgements}
\label{SectionV}
The authors thank Profs. V. Beskin, B. Somov and Yu. Lyubarsky for
useful discussions.
The work of the authors  is supported by RFBR grant 10-02-00599-a. MP also wants to acknowledge the support of  Presidential Grant of the Russian Federation to Support Young Russian
Candidates of Science (MK-1582.2010.2).
This research has
made use of NASA's Astrophysics Data System.


\bibliographystyle{spr-mp-nameyear-cnd}
\bibliography{burst}

\end{document}